# X-ray Coulomb explosion imaging reveals role of molecular structure in internal conversion

Till Jahnke[1,2,#], Sebastian Mai[3], Surjendu Bhattacharyya[4], Keyu Chen[4], Rebecca Boll[2], Maria Elena Castellani[5], Simon Dold[2], Avijit Duley[4], Ulrike Frühling[6], Alice E. Green[2,7], Markus Ilchen[2,6,13], Rebecca Ingle[8], Gregor Kastirke[9], Huynh Van Sa Lam[4], Fabiano Lever[6], Dennis Mayer[6], Tommaso Mazza[2], Terence Mullins[2], Yevheniy Ovcharenko[2], Björn Senfftleben[2], Florian Trinter[10], Atia Tul Noor[6], Sergey Usenko[2], Anbu Selvam Venkatachalam[4], Artem Rudenko[4], Daniel Rolles[4], Michael Meyer[4], Heide Ibrahim[11,12], and Markus Gühr[6,*]

[1]Max-Planck-Institut für Kernphysik, Saupfercheckweg 1, 69117 Heidelberg, Germany
[2]European XFEL, Holzkoppel 4, 22869 Schenefeld, Germany
[3]Institute of Theoretical Chemistry, University of Vienna, Wahringer Str. 17,1090 Vienna, Austria
[4]James R. Macdonald Laboratory, Kansas State University, Manhattan, Kansas 66506, USA
[5]University of Oxford and Rosalind Franklin Institute, Didcot OX11 0QS, United Kingdom
[6]Deutsches Elektronen-Synchrotron DESY, Notkestr. 85, 22607 Hamburg, Germany
[7]Stanford PULSE Institute, SLAC National Accelerator Laboratory, Menlo Park, California 94025, USA
[8]Department of Chemistry, University College London, 20 Gordon Street, London WC1H 0AJ, United Kingdom
[9]Goethe-Universität Frankfurt, Max-von-Laue-Str. 1, 60438 Frankfurt am Main, Germany
[10]Fritz-Haber-Institut der Max-Planck-Gesellschaft, Faradayweg 4-6, 14195 Berlin, Germany
[11]Advanced Laser Light Source @ INRS, Centre Énergie, Matériaux et Télécommunications, 1650 Boulevard Lionel-Boulet Varennes, Québec J3X 1P7, Canada
[12]Department of Physics, University of Ottawa, 150 Louis-Pasteur Pvt, Ottawa, Ontario K1N 6N5, Canada
[13]Department of Physics, Universität Hamburg, 22607 Hamburg, Germany

#till.jahnke@xfel.eu
*markus.guehr@desy.de

Surjendu Bhattacharyya: ORCID: 0000-0001-7107-8006
Rebecca Boll: ORCID: 0000-0001-6286-4064
Maria Elena Castellani: ORCID: 0000-0002-7868-2027
Alice E. Green: ORCID 0000-0002-6897-1247
Markus Gühr: ORCID 0000-0002-9111-8981
Heide Ibrahim: ORCID: 0000-0001-6371-8501
Markus Ilchen: ORCID: 0000-0001-5201-0495
Rebecca Ingle: ORCID: 0000-0002-0566-3407
Sebastian Mai: ORCID: 0000-0001-5327-8880
Huynh Van Sa Lam: ORCID: 0000-0003-4767-5390
Florian Trinter: ORCID: 0000-0002-0891-9180
Sergey Usenko: ORCID: 0000-0001-5202-5267
Anbu Selvam Venkatachala: ORCID: 0000-0002-5206-6955

# Abstract

Molecular photoabsorption results in an electronic excitation/ionization which couples to the rearrangement of the nuclei. The resulting intertwined change of nuclear and electronic degrees of freedom determines the conversion of photoenergy into other molecular energy forms. Nucleobases are excellent candidates for studying such dynamics, and great effort has been taken in the past to observe the electronic changes induced by the initial excitation in a time-resolved manner using ultrafast electron spectroscopy. The linked geometrical changes during nucleobase photorelaxation have so far not been observed directly in time-resolved experiments. Here, we present a study on a thionucleobase, where we extract comprehensive information on the molecular rearrangement using Coulomb explosion imaging. Our measurement links the extracted deplanarization of the molecular geometry to the previously studied temporal evolution of the electronic properties of the system. In particular, the protons of the exploded molecule are well-suited messengers carrying rich information on the molecule's geometry at distinct times after the initial electronic excitation. The combination of ultrashort laser pulses to trigger molecular dynamics, intense X-ray free-electron laser pulses for the explosion of the molecule, and multi-particle coincidence detection opens new avenues for time-resolved studies of complex molecules in the gas phase.

# Main

When molecules absorb light in the visible to ultraviolet range, a complex interplay of electronic and nuclear degrees of freedom converts light energy into other energy forms of molecular energy. Figure 1A depicts an exemplary sketch of such a light-driven molecular rearrangement. The molecule is excited from its ground state onto an electronically excited potential energy surface (PES) in the space spanned by the molecule's nuclear degrees of freedom. Following this electronic excitation, nuclear motion sets in along the gradient of the PES. The concept of a PES relies on a separation of fast electronic and slow nuclear motion due to their intrinsically different timescales, commonly referred to as Born-Oppenheimer approximation (BOA). During nuclear motion, the propagation on one PES can couple 'nonadiabatically' to another PES, if the two are energetically close to each other along the path in geometry space. This is the case, for example, for the cyan ($^1\pi\pi^*$) and blue ($^1n\pi^*$) PES in Fig. 1A along the shown exemplary reaction coordinate. This nonadiabatic coupling cannot be described in the framework of the BOA and requires the challenging treatment of coupled electron-nuclear dynamics [1]. In contrast to radiative decay, the nonadiabatic radiationless transitions are ultrafast and thus efficient reaction funnels. The prototypical topology for the coupling of different electronic states is a *conical intersection* (CoIn) of states (see, e.g., Refs. [2, 3]). Many processes in nature such as retinal light harvesting [4], optical switching of green and yellow fluorescent proteins [5, 6], and nucleobase photoprotection [7, 8] happen via nonadiabatic dynamics. In order to fully monitor and understand the nonadiabatic dynamics of a molecule, combined knowledge on aspects of two different realms is required. Namely, molecular geometry changes that drive the molecule towards regions of strong nonadiabatic coupling, and the resulting change of the electronic states. In the work presented here, we aim to precisely capture this full picture. We present a molecular movie with details about how the nuclear dynamics induce coupling among different electronic states. We provide unprecedented experimental details on the nuclear-motion pattern gathered from time-resolved Coulomb explosion imaging (CEI) [9] and combine these with measured changes of a molecule's electronic structure recorded in a previous experiment using time-resolved X-ray photoelectron spectroscopy (XPS) [10].

We studied the 2-thiouracil (2-tUra) molecule belonging to the group of thionated nucleobases. In contrast to canonical nucleobases, which undergo a fast radiationless transition into the electronic ground state after photoexcitation [7, 8], thionucleobases show an efficient population of long-lived triplet states after ultraviolet excitation [11, 12]. The efficient intersystem crossing is a hallmark of these molecules and has important consequences for possible applications. Some thiobases are used as medicinal drugs (e.g., for immunosuppression [13]) and can induce serious light-triggered damage in patients due to their triplet state induced photochemistry. The same photochemistry can be used, e.g., in photodynamic therapy and photocrosslinking studies [12, 14, 15]. Among the different thionucleobases, the microscopic origin of the efficient relaxation into the triplet state is best investigated in 2-tUra. Several simulations point at a mechanism that mostly occurs via the $S_1$ $^1n\pi^*$ state as a doorway state out of the optically excited $S_2$ $^1\pi\pi^*$ state [16, 17]. From that doorway state, the population further relaxes via El-Sayed allowed pathways to the $T_1$ $^3\pi\pi^*$ state. Simulations also suggest a distinct excited-state change in the molecular geometry, i.e., a breaking of the planar symmetry (see Fig. 1E1), as an underlying process. The relaxation out of the Franck-Condon (FC) region is suggested in simulations to occur under ring puckering at the $C_2$ atom, which is moving out of the molecular plane and thus induces a local pyramidalization (see Fig. 1E2). This motion is accompanied by an elongation of the C-S bond, which, however, occurs in the ring plane. The pyramidalization at $C_2$ is suggested to happen within the first 100 fs after the UV excitation and leads the molecule from the FC region over the $^1\pi\pi^*/^1n\pi^*$ CoIn towards the minimum of the $^1n\pi^*$ state [16, 17]. A breaking of the planar symmetry (Cs) is in general necessary for the molecule to change from $^1\pi\pi^*$ to $^1n\pi^*$ electronic character, even if the wave packet would propagate on one adiabatic state without undergoing nonadiabatic dynamics. Close to the $S_1$ $^1n\pi^*$ minimum (which the molecule reaches after the first 100 fs), the sulfur atom is suggested to point strongly out of the ring plane and resides in this configuration throughout the rest of the relaxation into the triplet states (see Fig. 1E3).

The UV-induced electronic relaxation of 2-tUra has been previously investigated employing time-resolved x-ray photoelectron spectroscopy at the FLASH free-electron laser studying sulfur 2p electrons [10]. Shifts in the time-resolved XPS indicate changes in the charge density, in analogy to the ground-state chemical shift [18, 19]. The passage from the $^1\pi\pi^*$ to the $^1n\pi^*$ state increases the UV-induced positive charge on sulfur leading to a photoline shift to lower kinetic energies by less than one eV up to 130 fs after the UV excitation. At this point, a maximal amount of molecular population has traversed the CoIn from $^1\pi\pi^*$ to $^1n\pi^*$. In addition, after 130 fs further oscillations in the XPS as well as in the $^1n\pi^*$ state population are visible, resulting from a repeated passage of the molecular wave packet through the CoIn. The results obtained by the time-resolved XPS are summarized in Fig. 1B and are shown with the simulated changes of the molecular geometry. We also show (Fig.1C) the simulated $^1n\pi^*$ population from Ref. [17] together with the experimental findings, indicating close agreement of the excited-state chemical shift (ESCS) with the molecular-dynamics simulations. While simulations based on the molecular geometry trajectories [16,17] explain the electronic dynamics, the geometric origin of the observed ESCS is not accessible in XPS, calling for an experimental study targeting explicitly these underlying geometry changes.

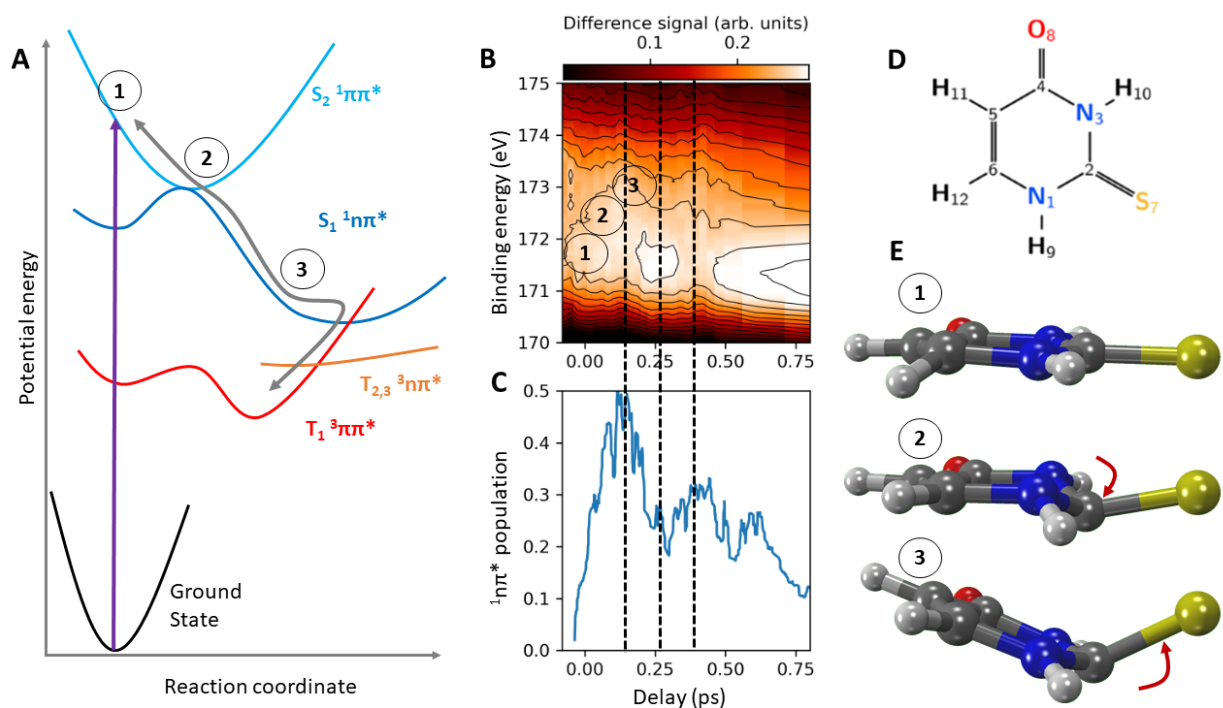

*Figure 1: **Summary of the 2-tUra molecular dynamics induced by UV excitation.** (A) Sketch of the molecular PES according to Ref. [16], showing the electronic states that are involved in the molecular dynamics from the optical excitation to being trapped in the triplet state. States are named by their order with the S, T symbols and their dominant electronic character. The Franck-Condon region, $S_2$-$S_1$ degeneracy and $S_1$ minimum are marked by points 1,2, and 3, respectively with suggested wave packet arrival times in (B) and simulated geometries in (E). Note that, in 2-tUra, the adiabatic $S_1$ state was found to be of $^1n\pi^*$ character across the relevant parts of the nuclear position space. (B) Time-resolved XPS together with (C) $^1n\pi^*$ population from quantum dynamics simulations [10,17]. (D) Molecular geometry and (E) predicted changes of the molecular geometry with UV excitation at positions 1,2,3 depicted in A - the symmetry-breaking motions are indicated by the red arrows.*

In our present work, we employ time-resolved CEI to examine the geometrical changes in the molecule, which underlie the previously observed temporal evolution of its electronic properties after UV excitation. CEI relies on a rapid charge-up of the molecule that causes its rapid fragmentation into (ideally) atomic ions. By measuring the three-dimensional momenta of these fragments, it is, in principle, possible to reconstruct the geometry of the molecule at the instant of ionization. Strictly speaking, this inversion from measured momentum space back to position space (without sophisticated molecular-dynamics modeling) is only possible for the ideal case of an instantaneous charge-up, as demonstrated by several examples [20-23]. Instead of an inversion, momentum and energy distributions from charge-up can be modeled and compared to the measured results [24]. CEI has been pioneered several decades ago, using first thin foils [25], and then short IR laser pulses for the charge-up [21, 26]. The latter is commonly used [23, 27-29]. This experimental avenue allowed to perform time-resolved CEI by using pump-probe schemes [30-34]. Smaller molecules were addressed relying on synchrotron radiation for the ionization of core electrons and subsequent (multiple) Auger-Meitner decay(s) generated the charge needed for the fragmentation [35, 36]. More recently, CEI work using (X-ray) free-electron lasers as a trigger for the fragmentation emerged [24, 37-41]. The usage of ultrashort/ultraintense pulsed X-rays has several advantages, e.g., typically more than one charge is generated per absorbed photon which helps to fulfill the requirement of a close to instantaneous charge-up of the molecule as well as reaching high charge states. Our experiment was performed at the Small Quantum Systems (SQS) instrument of the European X-ray free-electron laser (EuXFEL). UV laser pulses (266 nm) were used for electronic excitation of the 2-thiouracil molecules and intense X-ray pulses (h$\nu$ = 2.6 keV, duration 8-10 fs) were used for ionization. After the Coulomb explosion, the momenta of the generated ionic molecular fragments were measured in coincidence using the COLTRIMS reaction microscope [42, 43] at SQS. This technique allows targeting single molecules in the gas phase, providing the three-dimensional vector momenta of the charged particles generated by the ionization process. Please refer to the Methods section for details on the experimental setup, the properties of the UV and free-electron laser light, and the data analysis.

Figure 2 depicts molecular-frame momentum-space results of the Coulomb explosion of the molecule in its ground state. The molecular frame of reference is defined using the emission directions of the sulfur and oxygen ions. The measured $S^+$ momentum defines the $z$ axis and spans the $y$-$z$ plane together with the $O^+$ momentum. The momenta of all detected ions are then transformed into this coordinate frame by a rotation and are, in addition, normalized by the magnitude of the $S^+$ momentum. It has been shown that this normalization creates particularly clear momentum-space images when inspecting ring-like molecules [24]. Figure 2A shows the projection of these normalized momenta of the $O^+$ ions and protons on the $y$-$z$ plane. It depicts cleanly separated features corresponding to the four different hydrogen atoms of the molecule. In Fig. 2B, we plot the projection of the normalized proton momenta on the $x$-$z$ plane. Despite being a planar molecule, the protons exhibit a non-negligible contribution of out-of-plane emission, which probably corresponds to ground- or (thermally) excited-state vibrational motion in the unpumped initial state. While the sharp momentum-space features of the emitted protons in Fig. 2 indicate a rapid fragmentation of the molecule, the corresponding molecular-frame momenta of the carbon ions are smeared out, as we show in Extended data Fig. E1. Here, the X-ray-induced charge-up dynamics seem to prohibit an easy extraction of information on properties of the initial molecular geometry without referring to detailed modeling of the ionization process. The protons on the other hand are emitted very rapidly during the initial phase of the charge-up [24], thus carrying the desired information. Figure 2C depicts the three-dimensional angular emission distribution of the protons in a molecular frame defined by the difference and sum momenta of the $S^+$ and $O^+$ ions (see Methods section for details). The coordinate frame is depicted schematically at the left, including the definitions of the azimuthal and polar angles $\Phi$ and $\Theta$, respectively. The protons labeled 10 and 12 are located at the north and south poles, respectively. Proton 9 is located at an azimuthal angle of $\Phi = 0°$, proton 11 is located at $\Phi = +-180°$ (wrapping around). The S and O symbols mark the approximate emission direction of the sulfur and oxygen ions (in the planar configuration of the molecule).

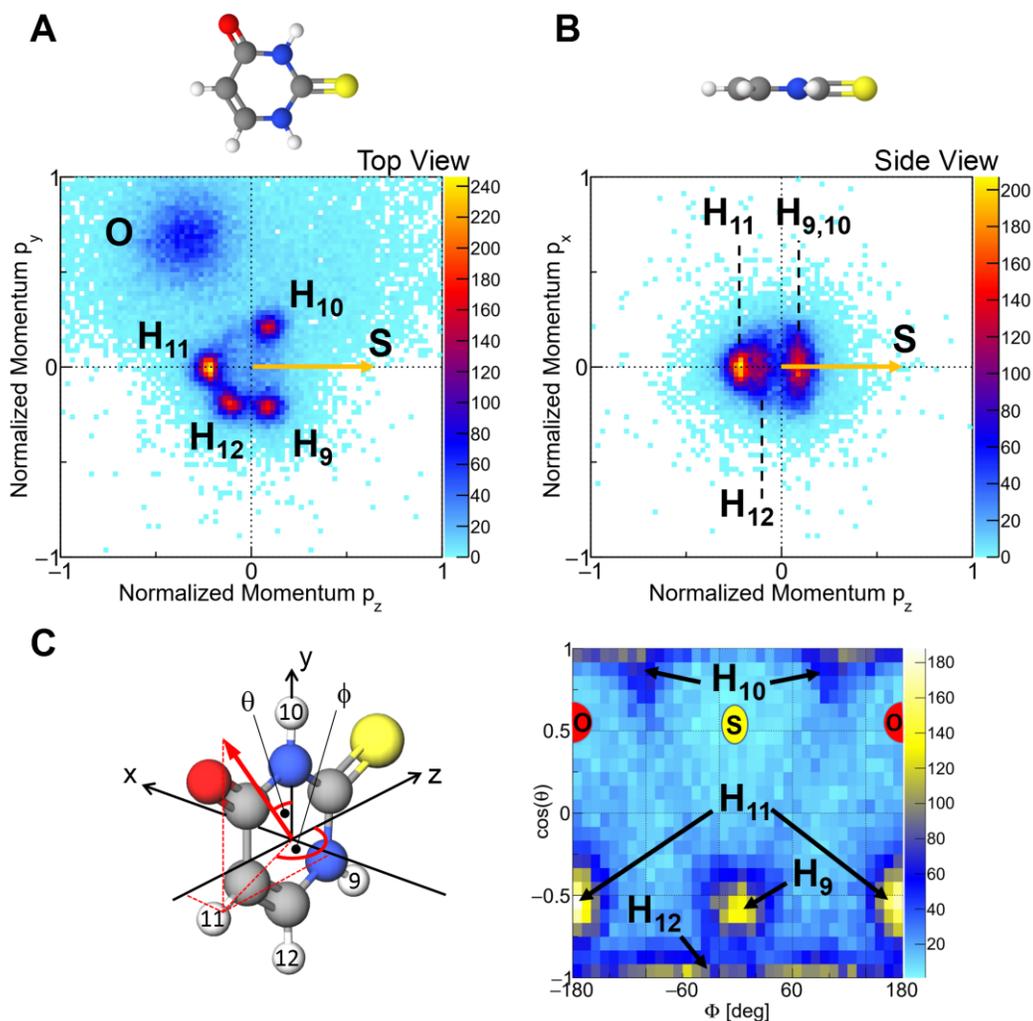

*Figure 2: **Coulomb explosion momentum-space images of 2-thiouracil without UV excitation.** The molecular coordinate frame is defined by the emission direction of the S atom providing the z axis. The y-z plane is spanned by the emission directions of the $S^+$ and $O^+$ ions. The x axis is perpendicular to the y-z plane. (A) "Top view" projection of the measured momentum-space data of the oxygen ions and the protons. (B) "Side view" projection of the measured proton momenta. The momenta have been normalized such that the magnitude of the sulfur momentum equals 1 for each molecular ionization event. (C) Angular emission distribution of the protons in the molecular frame indicated by the scheme at the left, which is defined by the difference and sum momenta of the $S^+$ and $O^+$ ions. In the given representation of the spherical coordinates $\Theta$ and $\Phi$, $H_{11}$ and O appear at two positions as the $\Phi$ angle wraps around. $H_{10}$ and $H_{12}$ are located at the poles and thus extend across a wide range of $\Phi$. For details see the main text.*

In order to show a UV-induced geometric deformation of the 2-thiouracil molecule (Fig. 1E), we focus on the relative emission direction of the sulfur and oxygen ions first. Theory predicts that the sulfur atom moves out of the molecular plane as time progresses (see Fig. 1). This rather drastic change of geometry should be observable already in the relative emission angle between the sulfur and the oxygen ions upon fragmentation by Coulomb explosion. In order to check this assumption, we refer to the modeled molecular dynamics (trajectory) data, initially presented in [16, 17]. We emulated the Coulomb explosion on this trajectory data by setting the charge of all atoms of the molecule to +1 at different timesteps of the molecular-dynamics simulation and then using our Coulomb explosion code to obtain the final-state momenta of all ions after fragmentation (see Methods section for details). Figure 3A shows the temporal evolution of the cosine of the relative emission angle $\alpha$ between the $S^+$ and $O^+$ ions as predicted by this modeling. At short times after the UV excitation, $cos(\alpha)$ is strongly peaked at a value of approximately -0.4. The distribution of $cos(\alpha)$ broadens for later delays with a more asymmetric contribution towards larger $cos(\alpha)$ values. In Fig. 3B, we show the measured $cos(\alpha)$ distribution for unpumped (blue) and UV-pumped molecules with a time delay of $\Delta t$ = 1000 fs (red). In line with the modeling, the experimentally determined distribution broadens and contributes to a larger extent on the side of higher $cos(\alpha)$ values. Panel C shows the evolution of $cos(\alpha)$ as difference plots between the UV-pumped and unpumped cases for four different pump-probe delays ($\Delta t$). The top row shows the results from trajectory modeling, the bottom row presents the experimental data. The measured distributions are wider compared to the ones obtained from theory, but most of the trends are present in the experiment. The main negative peak close to $cos(\alpha)$ = -0.4 becomes more pronounced over time, as does the asymmetry of the broadening of the $cos(\alpha)$ distribution that is observed in Figs. 3A and 3B. The negative dip in the experiment in Fig. 3C is much broader than in the simulation. We attribute this to an idealized modeling of the Coulomb explosion, not covering a finite charge-up time and neglecting the overall dynamics during the charge-up process. This results in a ground-state distribution in our simulated data which is narrower in $cos(\alpha)$ than the distribution measured in the experiment.

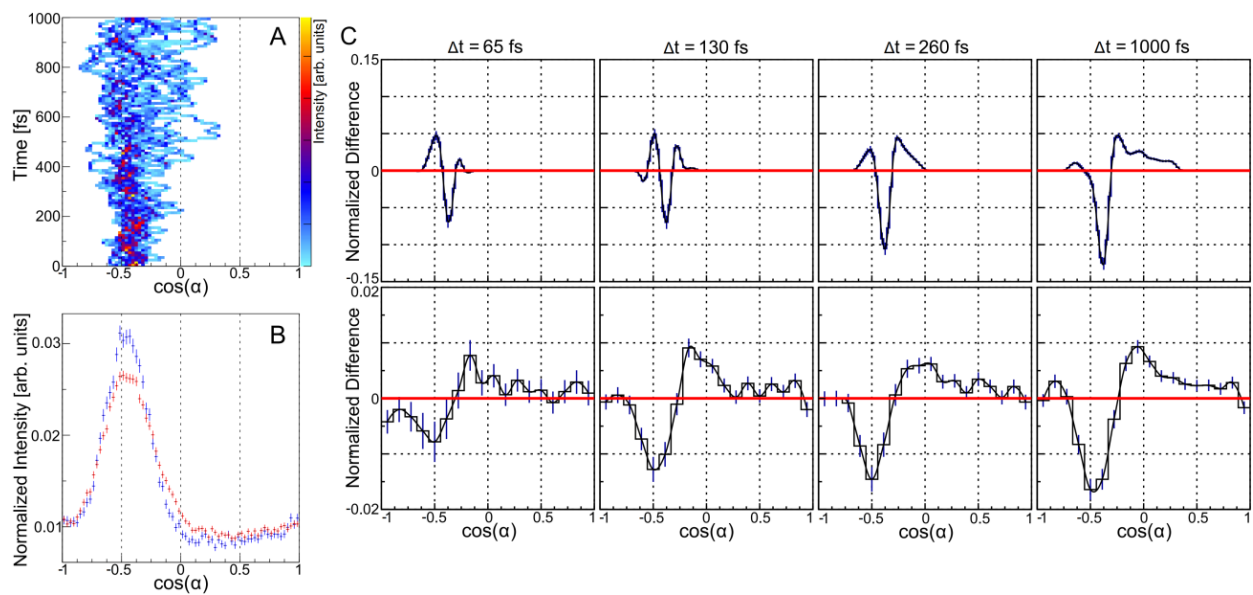

*Figure 3: **UV-induced changes of the relative emission angle α between the sulfur and the oxygen atoms.** (A) Temporal evolution of cos(α) predicted by our theoretical modeling. (B) Measured relative emission angle UV-pumped (red) and unpumped (blue). (C) Difference between UV-pumped and unpumped distributions of cos(α) for several time delays between the pump and the probe pulse. Top: modeling, bottom: measured results.*

To explore further details of UV-induced geometry changes, we now inspect the three-dimensional angular emission distributions of the protons shown in Fig. 2C as a function of the delay between the UV and the X-ray pulse. In Fig. 4, top row, we present the corresponding results obtained from our excited-state trajectory-based modeling. Note that the intensity (indicated by the color map) is normalized by the integrated number of ions per panel to visualize the relative intensity changes between different delay steps. At short times after the UV excitation, the proton emission directions are well-defined in the $\cos(\Theta)$-$\Phi$ space. As time evolves, firstly, these proton peaks broaden. Secondly, filament-like structures emerge in the range of $-0.1 < \cos(\Theta) < 0.6$ (yellow boxes). The bottom row depicts our measured results for the same time delays. The contrast is much weaker as compared to the top row, as only a part of the ground-state population in the experiments is UV-excited. Nevertheless, the loss of intensity in the peaks corresponding to $H_9$ (green boxes) and $H_{11}$ (red boxes) is clearly visible, as is an enhancement of the 'filament' structure at larger delays. The presence of traces of these filaments even at the earliest delays, although not as pronounced as at late times, may be due to vibrationally excited molecules (that are still in their electronic ground state), which occur as we heat our sample in the oven source in order to generate the gas-phase target.

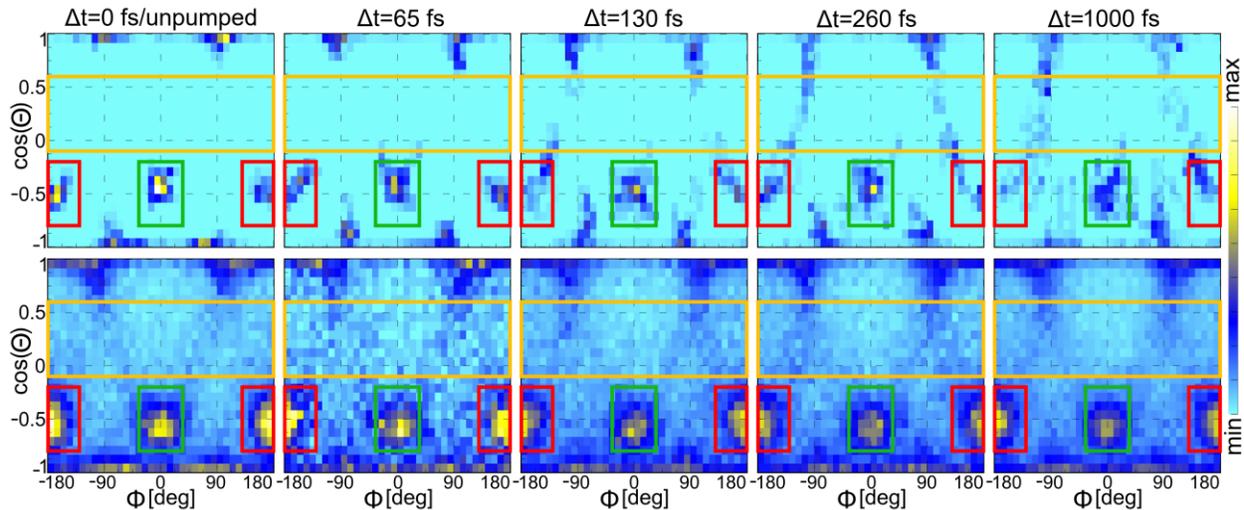

*Figure 4:* **Time-resolved proton emission direction in the spherical coordinates $\Phi$ and $\Theta$.** *Progression of the proton angular distributions for different pump-probe delays as provided by our quantum simulations (top row) [17] and measurements (bottom row). The signals at different delays are normalized to the integrated number of ions per delay step. The features associated with $H_9$ are in the green frame, $H_{11}$ is contained in the red frames. The so-called filaments are framed by the yellow box.*

As a final step, we integrated over the areas that are marked by three different colors in the panels of Fig. 4. Figure 5 shows how these integrals change for different pump-probe delays in theory (line) and experiment (dots with statistical error bars). In Fig. 5A, the integrated intensity belonging to the peak of the $H_9$ atom (green boxes in Fig. 4) is depicted, which diminishes in both the experiment and the simulation almost instantaneously, i.e., already 65 fs after the UV excitation. In comparison, the change of intensity of the $H_{11}$ peak (red boxes in Fig. 4, with the integrated signal shown in Fig. 5B) has a delayed onset, with the first decrease setting in at the next measured delay point of 130 fs after the UV pulse. In Fig. 5C, the increase observed in the "filament region" (yellow boxes in Fig. 4) starts with a slight delay, as well. We interpret these findings in the following way: within 65 fs after the UV excitation, the $H_9$ signal in the indicated green area diminishes and the angular emission distribution of $H_9$ broadens in the Φ-Θ plane. Clearly, the symmetry of the molecule is changing and the $H_9$ moves out of the plane spanned by the difference and sum of the oxygen and sulfur momenta. We interpret this as a consequence of the $N_1$-$C_2$-$N_3$ triangle and the attached protons moving out of the molecular plane as indicated by the theoretical prediction shown in Fig. 1E2. At later delays, the diminished $H_9$ signal stems from the sulfur atom moving out of the molecular plane (Fig. 1E3). In this geometry, all of the hydrogen atoms are located outside of the y-z plane, spanned by the oxygen and sulfur sum- and difference momenta. Thus, starting at 130 fs after UV excitation, also the $H_{11}$ signal in the red box diminishes due to the sulfur shifting out of the molecular ring plane. The difference in the onset of the decay of the $H_9$ and $H_{11}$ signals illustrates the differences between the onset of geometries (2) and (3) in Fig. 1E. The early onset of geometry (2) is the hallmark of a fast change from a planar to a nonplanar (puckered) geometry and in agreement with an immediate excited-state chemical shift in Fig. 1B. After 65 fs, we already observe that half of this shift is reached in the time-resolved XPS signal. This shows that the molecular pathway around the CoIn is described by the $C_2$ atom moving out of the plane according to geometry (2). Breaking the planar (Cs) symmetry is crucial for the nonadiabatic dynamics. First, the fast deformation of the molecule by a light atom (i.e., the $C_2$) determines the path of the molecular wave packet via the $^1\pi\pi^*$-$^1n\pi^*$ CoIn. Only after that, the mode including the heavier sulfur sets in. This is confirmed by the delayed onset of the corresponding feature in our CEI data, which thus agrees with the intuitive argument that motion of light atoms can induce nonadiabatic dynamics fast and thus efficiently [16].

In summary, our experiment allowed us to unravel the intertwined properties of the complementary electronic and nuclear molecular degrees of freedom. We obtained a direct view into the molecular dynamics using time-resolved X-ray CEI by using the emitted protons (which show well-defined features in momentum space) as messengers for details on the structural deformations of the molecule. The extraction of the information on the nuclear dynamics was possible without a model-based inversion of the recorded momentum-space data into real space. We observe the deplanarization of the molecule first at the $C_2$ atom at the earliest time after the excitation, which then continues after a short delay at the S atom. The deplanarization is at the origin of the nonadiabatic dynamics at the CoIn between the $^1\pi\pi^*$ and $^1n\pi^*$ state, and we are able to correlate the geometric properties of the molecule extracted from CEI with the electronic-state changes observed in X-ray photoelectron spectroscopy [10, 17]. The combination of both delivers a comprehensive picture of the important degrees of freedom in the coupled electronic and nuclear dynamics. As symmetry plays an important role in the mixing of molecular electronic states, the interpretation of the angular emission distributions of the ions after the Coulomb explosion is an ideal tool for future investigations of photoexcited dynamics of (biologically relevant) molecules. We anticipate that the reconstruction of the temporal changes of these angular distributions can be extended to more complex molecular systems. Possible candidates need to be prepared in the gas phase, and the size of these molecules is limited by the number of atomic ions that can be generated by single XFEL light pulses. However, with respect to the number of ions that needs to be detected in coincidence, our messenger-atom approach shows that already a threefold ion-coincidence measurement can reveal striking details on the UV-induced dynamics of a molecule as large as twelve atoms.

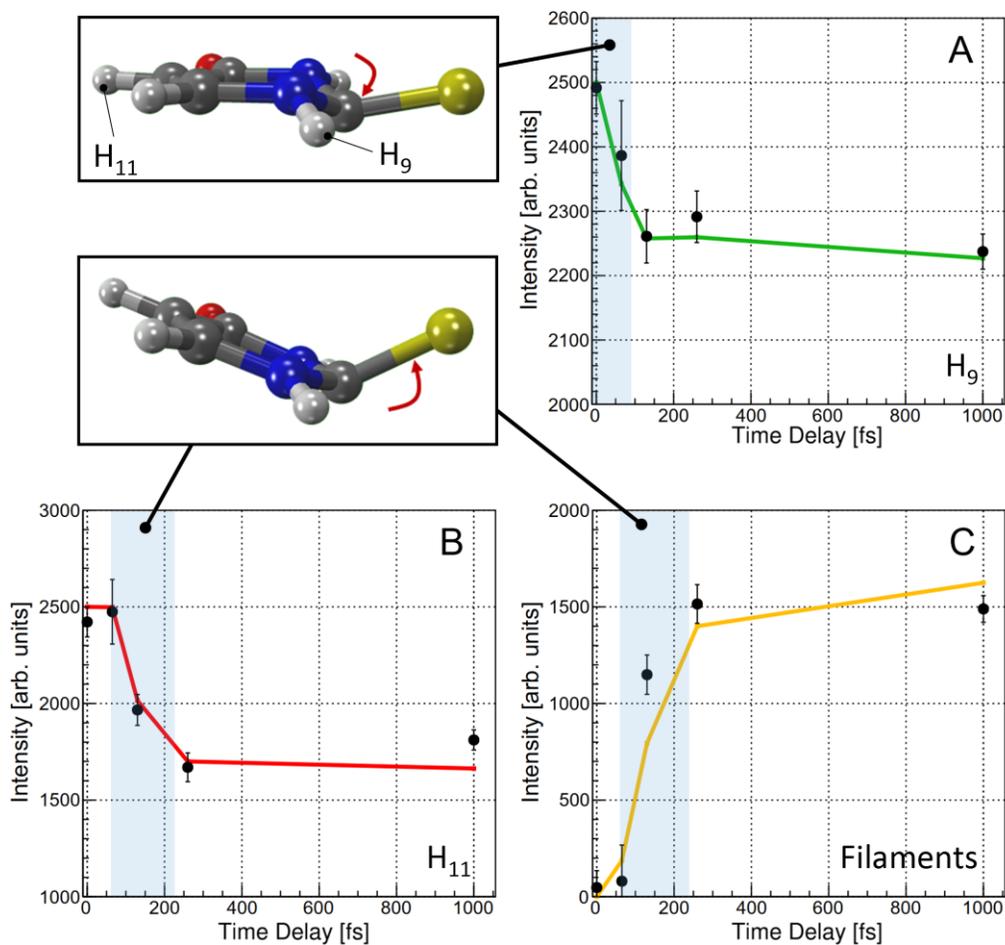

Figure 5: Integrated intensity extracted from Fig. 4, where the integration regions are marked by boxes of different colors. (A) and (B) angular emission regions of the $H_9$ and $H_{11}$ atoms as a function of the pump-probe delay. (C) Corresponding plot for the "filament region". The instant change of intensity in panel (A) corresponds to a change in the molecular geometry shown at the top left and the delayed onset in panels (B) and (C) to the one depicted at the middle left. The dots show the experimental results (with statistical error bars), the colored lines correspond to the outcome of our simulations. Note: the experimental results have been scaled, see text (and Methods section) for details.

# Methods

The experiment was performed at the SQS scientific instrument of the European X-ray Free-Electron Laser using cold target recoil ion momentum spectroscopy (COLTRIMS) [42, 43]. A state-of-the-art COLTRIMS reaction microscope is available at SQS as a permanent user endstation. In the following, we describe the experimental setup, the properties of the UV laser used for the excitation of the molecules and the X-rays employed for triggering the Coulomb explosion as well as details on the data analysis and the Coulomb explosion simulation. Details on the modeling of the molecular dynamics which yielded the trajectories used as an input to our Coulomb explosion simulation can be found elsewhere [16, 17].

## Experimental setup

We used the COLTRIMS technique to measure the momenta of several charged atomic fragments, which were generated in the Coulomb explosion of 2-thiouracil molecules, in coincidence. At room temperature, the target molecules exist in the form of a powder. Accordingly, in order to generate a supersonic gas jet consisting of 2-thiouracil molecules, we evaporated the substance in an oven, which was part of a nozzle generating the gas jet. The oven was held at a constant temperature of 240 °C. The supersonic jet formed as a mixture of the vapor and He carrier gas (which was applied with a stagnation pressure of 0.5 bar to the oven) expanded through the nozzle hole (i.e., an aperture with a diameter of 200 μm) into vacuum. Before entering the main chamber housing the COLTRIMS analyzer, the gas jet was skimmed/collimated in three stages to form a well-localized target. The final collimation stage is typically used to reduce the amount of target such that the coincident measurement of ions originating from a single molecule can be performed. The target beam was crossed at right angle with the light from the XFEL and the exciting UV laser. Ions which were created in the target region upon the interaction with the ionizing light were guided by an electric spectrometer field to a time- and position-sensitive microchannel-plate detector with hexagonal delay-line position readout and an active diameter of 120 mm. The spectrometer consisted of an acceleration region (covering the interaction volume of the gas jet and the light beams) with a length of 60 mm and a strong electric extraction field of 330 V/cm applied. The acceleration region was followed by a drift region (with a length of 120 mm), which was passed by the ions before they hit the detector. The ions detected with respect to each XFEL pulse were recorded in coincidence. We reconstructed the trajectory of each individual ion inside the spectrometer from its flight time and the position of impact in an offline analysis and deduced from this information the mass-over-charge ratio of the ion and its initial momentum vector after the Coulomb explosion.

## Properties of the UV laser and the X-rays

The European X-ray Free-Electron Laser provides trains of short X-ray pulses at a repetition rate of 10 Hz with train/burst duration of up to 600 μs. It was operated in a mode which yielded a maximum rate of pulses within such a train of 1.1 MHz. In order to incorporate the flight time of heavy molecular fragment ions, we employed every sixth of these pulses for our experiment yielding approximately 80 XFEL pulses per train in our measurement (and thus an effective repetition rate of 800 Hz). As a photon energy for the Coulomb explosion, we chose h$\nu$ = 2.6 keV, well above the sulfur K-edge, which is situated at 2472 eV for atomic sulfur [44]. We obtained a single-shot pulse energy of approximately 2 mJ, which was measured by a gas monitor detector upstream of the beamline. The transmission of the beamline was estimated to be 85% for this range of photon energies and the given configuration [45], which yielded a pulse energy of approximately 1.7 mJ on target. The focus diameter of the X-ray beam at the interaction volume was tuned by looking at the high-charge yield from multiphoton ionization of Xe atoms inspecting the corresponding time-of-flight spectra and estimated as 5 µm FWHM based on an off-line microscopy characterization under the same conditions. The X-ray pulse length was about 8-10 fs determined indirectly by analyzing the spectral distribution of the SASE pulses [46].

The third harmonic of an optical laser operating at 800 nm and synchronized to the X-ray pulses was used for the excitation of the molecules [47]. Pulses with energies up to 8.5 µJ of 266 nm radiation were delivered to the interaction volume. The duration of the UV pulses was measured to be about 35 fs and the focus was set to a diameter of approximately 120 µm. The molecular absorption cross section is in the range of 30 Mbarn [48], leading to a saturation of the $S_0$-$S_2$ molecular transition in the center of the UV focus. We note that two-photon ionization by the UV beam does excite a part of the population in the cationic ground state from the one-photon excited $S_2$ state. However, this does not interfere with the observation of deplanarization in the excited state, as the cationic ground state of 2-thiouracil is expected to remain planar. In this cationic state, the π* orbital localized on the $C_2$ atom that leads to deplanarization is not occupied [49]. The planar cationic ground state has been confirmed by an optimization at the MS(2)-CASPT2(11,9)/cc-pVDZ level of theory.

The temporal delay and the jitter between the X-ray and the UV pulses were monitored on a shot-by-shot basis using a pulse arrival-time monitor [50] which allows for time-resolved measurements with an overall temporal resolution of better than 40 fs.

## Data analysis

As indicated above, molecular fragment ions originating from a single molecule were detected in coincidence and their vector momenta were obtained from their impact positions on the detector and their flight times after the photoreaction. The mass-to-charge ratio of the measured ions is determined from their measured flight times. Then the full momentum information is retrieved from the recorded data by reconstructing the ions' trajectories inside the COLTRIMS spectrometer for each individual ion. From the momenta, all derived quantities such as ion kinetic energies and emission angles are deduced. In particular, due to the coincident detection of the ions, relative emission angles between different ions are obtained. This advantage of a COLTRIMS measurement is used to define two different molecular coordinate frames, which are used to present the data. Firstly, the momentum of the $S^+$ ion acts as the *z* axis of the coordinate frame, and the *y-z* plane is spanned by taking in addition the momentum of the $O^+$ ion into account. By definition, the *y* component of the $O^+$ momentum is greater than or equal to zero. The *x* axis is perpendicular to the *y-z* plane. The laboratory-frame momenta are then transformed into this molecular frame for each individual molecule which was exploded by the XFEL. In addition to this coordinate transformation, all momenta were scaled by the magnitude of the S$^+$ momentum prior to plotting. Secondly, in order to obtain the angular emission distributions shown in Figs. 2 and 4, a similar molecular coordinate frame is employed, which uses instead of the sulfur and oxygen ion momenta the sum and difference of these two momenta as a reference. Within this (much less intuitive) coordinate frame, the molecule is approximately oriented as indicated in Fig. 2 (for cases where it is in its planar ground-state equilibrium geometry).

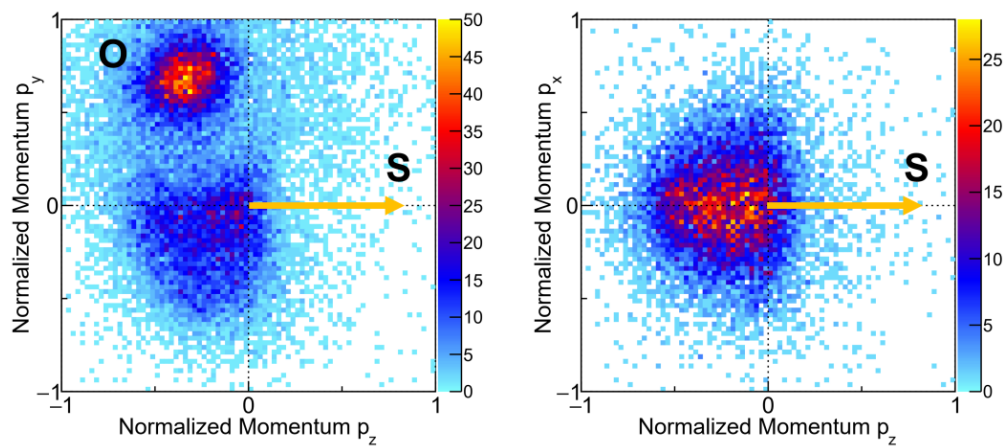

*Figure E1:* **Coulomb explosion momentum-space images of 2-thiouracil without UV excitation: Carbon ions.** *The molecular coordinate frame is defined by the emission direction of the S atom providing the z axis. The y-z plane is spanned by the emission directions of the S and O atoms. The x axis is perpendicular to the y-z plane. Left: "Top view" projection of the measured momentum-space data of the oxygen atom and the carbon ions. Right: "Side view" projection of the measured carbon ion momenta. The momenta have been normalized such that the magnitude of the sulfur momentum equals 1 for each molecular ionization event.*

The data presented in this article consist of three-fold coincidences, i.e., a coincident detection of an S$^+$, an O$^+$, and a H$^+$ ion. In line with previous results obtained at the EuXFEL [24], we conclude that the rapid charge-up triggered by the XFEL occurs along a well-defined route resulting in Coulomb explosion patterns which are (in particular) stable on a shot-to-shot basis. Therefore, a full picture of the Coulomb explosion is obtained even though only a subset of all generated ions is detected for each XFEL shot. Other than reported earlier [24], the charge-up observed in the present experiment was, however, less well-defined. We, therefore, rely in the main part of the article on the momentum-space Coulomb explosion patterns of the protons (inspected in a coordinate frame provided by the momenta of the O and S ions). The protons show the well-defined distributions visible in Fig. 2. The corresponding Newton diagrams showing the C$^+$ momenta in the same molecular coordinate frame are depicted in Fig. E1 and show only a rather broad and comparably featureless distribution. Our work, therefore, highlights the possibility to employ the protons (which are emitted very rapidly at the beginning of the charge-up by the XFEL) as snapshot-like messengers conveying details on the molecular geometry even in cases where the full charge-up does not yield a picture that is easy to interpret without employing sophisticated theoretical modeling.

In the experiment, we recorded separate datasets at pump-probe delays of 65 fs, 130 fs, 260 fs and 1 ps and without UV pump (i.e., X-rays only). We recorded a similar amount of statistics for each delay step, with the exception of the dataset belonging to a pump-probe delay of 65 fs, which has less statistics (as indicated, e.g., by the corresponding statistical error bar in Fig. 5). In order to compare the different datasets, we normalized the measured data by the integrated number of shots recorded for each set. In Fig. 5, we integrated over certain regions of the angular emission distributions of the protons as obtained from our measurements and our Coulomb explosion simulation. The absolute values retrieved from this integration cannot be compared between the simulation and the experiment. In addition, other than in the simulation, there is a strong contribution from unpumped molecules in the measured distribution. Accordingly, only the shape of the time dependence of the integrated intensities can be compared between experiment and simulation, and the experimental data points have been scaled with respect to the simulation. The error bars provided in Fig. 5 correspond to the statistical error of each integrated value after being scaled accordingly.

## Simulated Coulomb explosion

We employed the trajectory data published earlier [17] as an input for a simple Coulomb explosion model in order to compare the resulting momentum-space information with our measured time-resolved Coulomb explosion results. The used trajectory data covered the evolution of the molecule for a time of 1 ps after the UV excitation in time steps of 5 fs. Our Coulomb explosion code generates the momentum-space distribution of all atoms of the molecule by solving Newton's equations of motion for each time step of the provided trajectories after initializing each atom of the molecule at the given geometry with a charge of 1. Accordingly, our CE model assumes an instantaneous Coulomb explosion of the molecule into singly charged atomic fragments with purely Coulombic interactions. As the real Coulomb explosion is not instantaneous, a main difference between the modeled and the measured results is (typically) the overestimation of the kinetic energy of the ions after the Coulomb explosion. This discrepancy vanishes mostly when considering normalized momenta as depicted in Fig. 2. The angular emission pattern is affected by the model assumptions, as well, but (as the results presented in Fig. 4 indicate) to a lesser extent than the final-state energies. Figure E2 shows the temporal evolution of the proton and oxygen ion momenta as obtained from the Coulomb explosion simulation in the form of Newton plots (same representation as in Fig. 2A). Figure E3 shows the simulated proton momenta in correspondence to Fig. 2B.

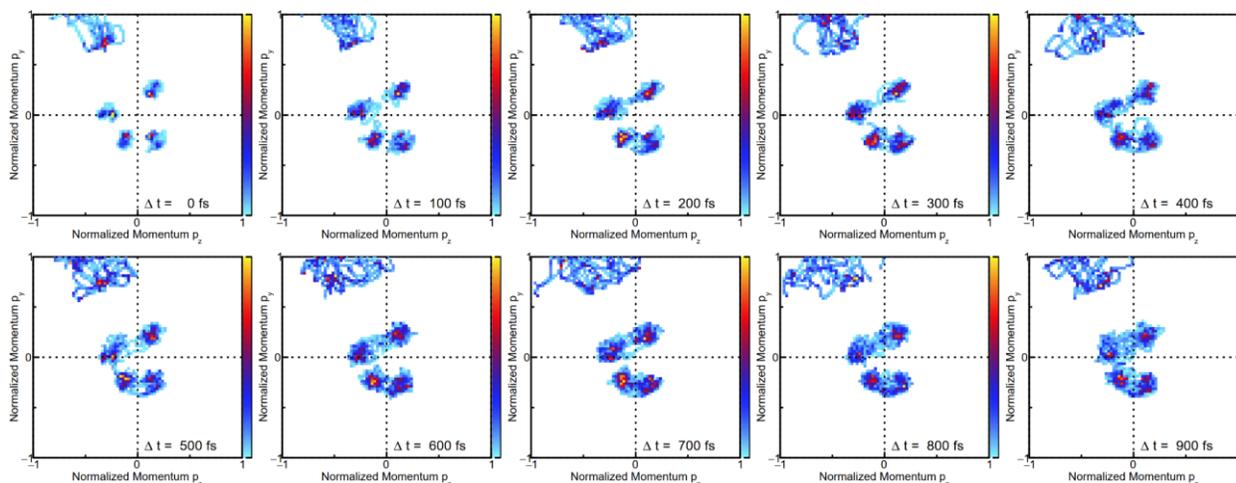

*Figure E2: **Time-dependent Newton plots showing simulated data, y-z plane.** Results from our Coulomb explosion simulation for a sequence of time steps (indicated in each panel at the bottom, right) as Newton plots (i.e., in the same representation used for the measured results in Fig. 2A).*

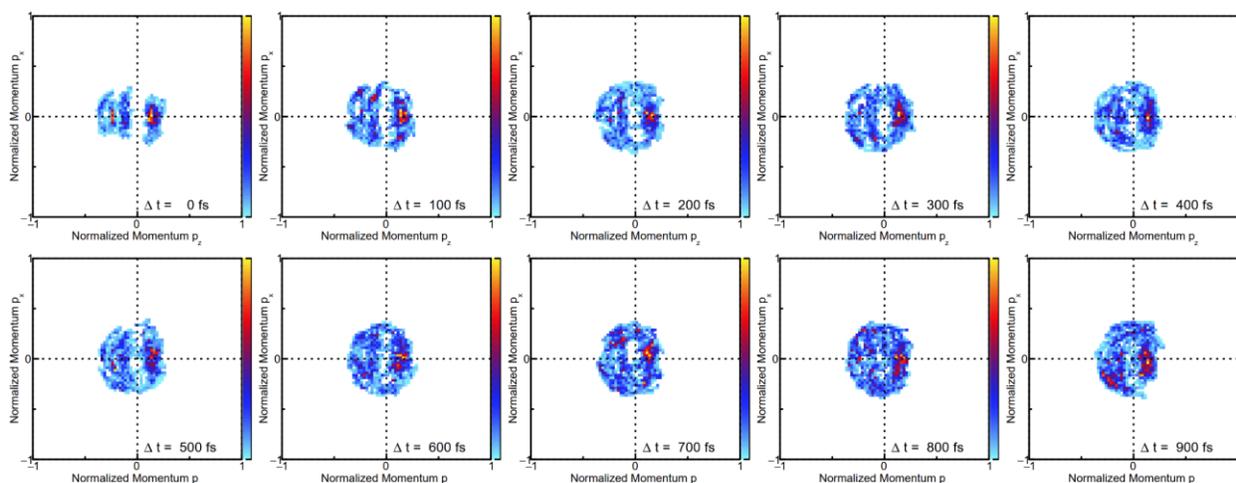

*Figure E3: **Time-dependent Newton plots showing simulated data, x-z plane.** Results from our Coulomb explosion simulation for a sequence of time steps (indicated in each panel at the bottom, right) as Newton plots (i.e., in the same representation used for the measured results in Fig. 2B).*

**Acknowledgements**

We acknowledge European XFEL in Schenefeld, Germany for the provision of XFEL beam time at the SQS instrument and would like to thank the staff for their assistance. Data recorded for the experiment at the European XFEL are available at https://doi.org/10.22003/XFEL.EU-DATA-003155-00.

**Author contributions**

TJ, SB, KC, RB, MEC, SD, HVSL, UF, AEG, MI, RI, GK, FL, TMa, TMu, YO, BS, ATN, SU, ASV, AT, DR, MM, HI, and MG performed the experiment. TJ analyzed the experimental data and the Coulomb explosion simulation results. SM simulated molecular trajectories. SB, KC, HVSL, ASV, AR, and DR provided the Coulomb explosion code and performed the Coulomb explosion of the simulated molecular trajectories. TJ, HI, and MG discussed the data and simulations and worked on an interpretation. TJ, SM, HI, and MG produced the figures and wrote the first draft of the paper. All authors contributed to the final version of the paper in iterative discussions.

**Funding**

HVSL, AR, and DR were supported by the Chemical Sciences, Geosciences, and Biosciences Division, Office of Basic Energy Sciences, Office of Science, US Department of Energy under grant no. DE-FG02-86ER13491. SB and KC were supported by grant no. DE-SC0020276 from the same funding agency, and AD and ASV by grant no. PHYS-1753324 from the National Science Foundation. FT acknowledges funding by the Deutsche Forschungsgemeinschaft (DFG, German Research Foundation) - Project 509471550, Emmy Noether Programme. HI acknowledges the Natural Sciences and Engineering Research Council of Canada and the NRC Quantum Sensors Project. M.M. acknowledges support by the Cluster of Excellence 'Advanced Imaging of Matter' of the DFG—EXC 2056 and project ID 390715994. AEG acknowledge funding by the European Union under project 101067645. SM acknowledges funding from the Austrian Science Fund (FWF), grant DOI 10.55776/P25827.

**Competing interests**

The authors declare no competing interest.